# Lambda, the Fifth Foundational Constant Considered by Einstein

Gilles Cohen-Tannoudji

Laboratoire de recherche sur les sciences de la matière (IRFU, CEA Université Paris-Saclay)

**Abstract** The cosmological constant, usually named Lambda, was introduced by Einstein in 1917 and then abandoned by him as his biggest "blunder". It currently seems to make a spectacular comeback in the framework of the new cosmological standard model. One will explain why, together with the Planck's constant, the Boltzmann's constant, the celerity of light and the Newton's constant, also considered by Einstein, the cosmological constant may play a foundational role in the conceptual framework, and in the metrological framework a role comparable with the one attributed to the Avogadro constant.

## 1/ Introduction

When I was invited to participate to the symposium devoted to the role of some fundamental constants in the redefinition of the international system of units (SI), I was asked to comment on the role of the fundamental constants as seen by Einstein and to try and explain what is for Einstein (and for me) the status and the profound meaning of the five constants that will redefine the SI: $h, k, c, e, N_A$. I realized that Einstein also considered five constants:

- the first three of them, $h, k, c,$ coincide with the ones considered for the revision of the SI;
- instead of $e$, he considered $G$ (the Newton's gravitational constant);
- and, instead of the Avogadro's constant $N_A$, he considered the cosmological constant $\Lambda$.

My intent, in this paper is to show that the purpose of Einstein, one century ago was not far from the purpose of the SI community and that, somehow, the cosmological constant can be related to an "Avogadro constant of spacetime".

In the second section, I will review the interpretation by Einstein of the dual role played by the universal constants: as conversion factors in a metrological framework and as constants reflecting some foundational principles. In the third section, I will focus on the cosmological constant $\Lambda$, in relation with the principle of the relativity of inertia which is at the foundation of the general theory of relativity. Einstein introduced $\Lambda$ in 1917 and then abandoned it qualifying it as his biggest blunder. Now, $\Lambda$ currently seems to be making a spectacular comeback. In the fourth section, I will explain why I believe that the five constants considered by Einstein play a foundational role, both in the metrological and conceptual frameworks.

## 2/ Einstein and the universal constants

### *2.1 Terminology and classification of the constants*

Since the terminology about the constants depends strongly on their interpretation, it may be useful to make precise the terminology I am going to use according to my own interpretation. I tend to distinguish two categories of constants:

1. The ones that I call *universal foundational physical constants,* that are *universal* because they are implied in the whole of physics and *foundational* (I prefer this word to *fundamental*) because they underlie the axiomatic conceptual framework and the basics units of physics
2. The other constants, such as *e* say, that are dimensionless once the units have been redefined thanks to the constants of the first category, are *parameters* in the framework either of

    *models,* such as the standard model of particle physics, in which case they must be determined experimentally, or of

    *theoretical conjectures* that would, possibly, allow calculating them.

In principle, all these constants are supposed to be really constant. Any well confirmed variation of anyone of them would imply either the existence of some physics *beyond the standard models* or the questioning of the general axioms of theoretical physics, a feature that enhances the stakes of a reliable most accurate metrology.

### *2.2 The apogee of classical physics*

At the end of the 19th century, the apogee of classical physics, the culmination of the scientific revolution which had seen the rise and the development of modern science, consisted of several theories that succeeded into three important syntheses or unifications and which allowed to model in a satisfactory manner all the phenomena that were then observable:

- The *electromagnetic theory of light* of Faraday, Maxwell and Hertz who unified the electrical, magnetic and optical phenomena,
- The *theory of universal gravitation* of Galileo and Newton who unified terrestrial and celestial mechanics and

- The upgrading of the conceptual status of the *atomistic conception* to the one of a genuine scientific discipline by means of the *molecular theory of matter* and the *statistical thermodynamics* of Maxwell and Boltzmann.

The successes achieved by these theories were such that Lord Kelvin (William Thomson), analyzing in 1900 the field of investigation of physics, announced that it was in the process of completion except for two "small clouds", of which he thought they would only require some adjustments to be resorbed. It was about the failure of the detection of the movement of Earth in the ether (experiment of Michelson and Morley) and the absence of a theoretical explanation to the observed spectrum of the black body, problems to which were added the one of the *photoelectric effect* and the one of the *precession of the perihelion of Mercury*.

### *2.3 The scientific revolution of the twentieth century*

It is well known that the considering, in particular by Einstein, of four universal constants, the constant of gravitation $G$, the speed of light in vacuum $c$, the Planck's constant $h$ and the Boltzmann's constant $k$ is at the origin of the relativity and quantum theories, thanks to which the "small clouds" have been dissipated, and which are at the basis of the scientific revolution of the twentieth century. In this respect, these four constants that belong to the first group of constants defined above, have played a foundational role.

The structure in tripod of the theoretical framework of classical physics is still convenient for the physics of the twentieth century. This framework now includes three theories each considering two of these four universal constants and extending the theories of the framework of classical physics:

- The *quantum theory of fields* (constants $h$ and $c$) extends, encompasses the electromagnetic theory of light and restores it as a quasi-classical approximation; it is the basis of the high energy physics which explores, at sub-atomic or even sub-hadronic scales, the structure of matter (elementary constituents and fundamental interactions). The discovery in 2012 at CERN, with the LHC, of the Higgs boson, keystone of the standard model (SM) of this physics, is the mark of a true apogee.
- The theory of *general relativity* (constants $G$ and $c$) extends, encompasses and restores as a nonrelativistic approximation the theory of universal gravitation of Newton. It is at the foundation of modern cosmology, which, since the first model of the Big Bang taking account of the expansion of the universe at extragalactic scales, up to the current apogee of observational astrophysics, is now also equipped with a genuine cosmological standard model (CSM).

- *Quantum Statistics* (constants $h$ and $k$) extends the analytic mechanics, the kinetic theory of matter and the statistical thermodynamics and serves as the basis for modeling (or for numerically simulating) all phenomena which, from the scale of the atom to that of the galaxy, rely on statistical physics and about which the quantum effects cannot be neglected. This physics is the framework of the phenomenological consolidation of the standard models of the physics of the structure of matter (SM) and of the cosmology (CSM), in terms of a *scientific cosmogony* the purpose of which, according to Georges Lemaître, "is to search for ideally simple initial conditions from which can result, by the interplay of known physical forces, the current world in its full complexity [1]. " Whereas general relativity and quantum theory of fields form the basis for what could be called the physical laws and fundamental structures of the universe, quantum statistics forms the basis of the physics of the *emergence* of the observable world at our scales. The role of this third component of the theoretical framework that considers not only the Planck's $h$ constant but also the Boltzmann constant $k$ is, in my opinion, not less important than that of the first two ones: it will be shown below, in section 5, that, as a consequence of the *informational turn of the interpretation of quantum physics*, this role is rather *foundational* in the sense that it fixes the conditions of the possibility of physics, among them, the existence of a reliable metrology, which, in my opinion, is exactly the aim of the redefinition of the SI thanks to the considering of universal constants.

## 3/ A brief history of the cosmological constant

### 3.1 At the onset of relativistic cosmology, the Einstein/de Sitter debate over the cosmological constant and the principle of the relativity of inertia

The reason why Einstein, in 1916, became interested in cosmology was that he wanted to clarify the relation between the theory of general relativity he was establishing and what he called the Mach's principle. This relation is well explicated in one of the conferences he gave in Princeton [2], in 1921:

> "The assumption of the complete physical equivalence of the systems of coordinates, K and K' we call the " principle of equivalence"; this principle is evidently intimately connected with the theorem of the equality between the inert and the gravitational mass and signifies an extension of the principle of relativity to co-ordinate systems which are in non-uniform motion relatively to each other. In fact, through this conception we arrive at the unity of the nature of inertia and gravitation. For according to our way of looking at it, the same masses may appear to be either under the action of inertia alone (with respect to K) or under the combined action of inertia and gravitation (with respect to K'). The possibility of explaining the numerical equality of inertia and gravitation by the unity of their nature gives to the general theory of relativity, according to my conviction, such a superiority over the conceptions of classical mechanics, that all the difficulties encountered in its development must be considered as small in comparison. (…) Although all of these effects are inaccessible to experiment, because κ is so small, nevertheless they certainly exist according to the general theory of relativity. We must see in them a strong support for *Mach's ideas as to the relativity of all inertial actions*. If we think these ideas consistently through to the end we must expect the *whole* inertia, that is, the whole $g_{\mu\nu}$-field, to be determined by the matter of the universe, and not mainly by the boundary conditions at infinity."

From this quotation, we can understand why, on the one hand, Einstein got interested in cosmology – because according to the "Mach's ideas", inertia of a test body is but the gravitation exerted on it by the rest of the universe, and on the other hand, why he considered a *finite universe*[1] – because he wanted to avoid uncontrollable assumptions about the boundary conditions at infinity.

But, with a finite universe he faced the problem of a universe which should eventually collapse under the action of its own gravitation; he thus tried a model involving a static universe [3], thanks to a *cosmological term* equal to the cosmological constant Λ multiplying the metric field, that in principle can be added in the left-hand side of his equation and that he had previously discarded.

The controversy between Einstein and de Sitter was about the Mach's principle that they agreed to name the *postulate of the relativity of inertia*. This controversy is accurately exposed in the three papers that de Sitter published in 1916-1917 [4]. In the first cosmological model that Einstein had proposed in 1917[3], he had enunciated the principle of the relativity of inertia to which he refers as the Mach's principle

"In any coherent theory of relativity, there cannot be inertia with respect to 'space' but only inertia of masses with respect to one another. Consequently, if in space, I take a mass far enough from all the other masses in the universe, its inertia must go to zero".

---

[1] Actually, as noted by Cormac O'Raifeartaigh (whom I acknowledge for his remarks), in the original version of [3], in German, Einstein refers to his model as 'spatially closed' rather than 'finite'

In the correspondence, they had in March 1917, Einstein and de Sitter agreed on a formulation of this principle which makes of it a genuine foundational principle: in a postscript added by de Sitter at the end of his second paper of ref. [4] he refers to and endorses a statement made (in German) by Einstein:

### Postscript

Prof. Einstein, to whom I had communicated the principal contents of this paper, writes "to my opinion, that it would be possible to think of a universe without matter is unsatisfactory. On the contrary the field $g_{\mu\nu}$ *must be determined by matter, without which it cannot exist* [underlined by de Sitter] This is the core of what I mean by the postulate of the relativity of inertia". He therefore postulates what I called above the logical impossibility of supposing matter not to exist. I can call this the "material postulate" of the relativity of inertia. This can only be satisfied by choosing the system A, with its world-matter, i.e. by introducing the constant $\lambda^2$, and assigning to the time a separate position amongst the four coordinates.

On the other hand, we have the "mathematical postulate" of the relativity of inertia, i.e. the postulate that the $g_{\mu\nu}$ shall be invariant at infinity. This postulate, which, as has already been pointed out above, has no physical meaning, makes no mention of matter. It can be satisfied by choosing the system B, without a world-matter, and with complete relativity of the time. But here also we need the constant $\Lambda$. The introduction of this constant can only be avoided by abandoning the postulate of the relativity of inertia altogether.

In this postscript, de Sitter also summarizes all the issues of the debate he had with Einstein and which are discussed in the core of his three papers. What he calls the "system A" refers to the Einstein's cosmological model of [3], i.e. a spatially finite universe obeying the Einstein's equation to which has been added a cosmological term (the "constant $\Lambda$") allowing to satisfy the "material principle of the relativity of inertia" thus playing the role of a hypothetical matter, "of which the total mass is so enormously great, that compared with it all matter known to us is utterly negligible[3] This hypothetical matter I will call the *world-matter*".

The principle of the relativity of inertia is indeed related to the Mach's principle:

"To the question: If all matter is supposed not to exist, with the exception of one material point which is to be used as a test-body, has then this test-body inertia or not? The school of Mach requires the answer *No*. Our experience however decidedly gives the answer *Yes*, if by 'all matter' is meant all ordinary physical matter: stars, nebulae, clusters, etc. The followers of Mach are thus compelled to assume the existence of still more matter: the 'world-matter'. If we place ourselves on this point of view, we must necessarily adopt the system A, which is the only one that admits a world matter."

---

[2] In the Einstein's and de Sitter's articles, the cosmological constant was written as lower case λ, in modern cosmology it is usually written as an upper case $\Lambda$ which is the convention we shall adopt in the present paper

[3] It is interesting to note that, already in 1917, it was realized that the known matter represents only a negligible part of the content of the universe.

The last statement of this quotation is reinforced in the postscript in which the world-matter of system A is identified with the constant λ.

To the "system A" is opposed the "system B" which is the well-known de Sitter universe containing no matter ($\rho = 0$) and which, nevertheless is a solution to the Einstein's equation with a cosmological constant. He exposed to Einstein this solution in a letter dated on March 20th and received on March 24th the Einstein's answer that he commented in the above quoted Postscript added to his communication on March 31st in front of the KNAW. The reproach made by Einstein to this "system B" solution of de Sitter was based on a three-fold argument: i) the corresponding universe is spatially finite, ii) it is bounded by a singularity[4], and iii) this singularity is at finite distance. In return, de Sitter made about the "system A" solution of Einstein the very severe criticism that it does not satisfy complete time relativity, but he had to recognize that his "system B" solution satisfies only a "mathematical" principle of relativity of inertia that he formulates in the following way:

> Once the system of reference of space- and time-variables has been chosen, the Einstein' equations determine the $g_{\mu\nu}$ apart from *constants of integration* [underlined by me], or boundary conditions. Only the deviations of the actual $g_{\mu\nu}$ from these values at infinity are thus due to the effect of matter. (…) If at infinity all $g_{\mu\nu}$ were *zero*, we could truly say that the whole of inertia, as well as gravitation, is thus produced. This is the reasoning which has led to the postulate that at infinity all $g_{\mu\nu}$ shall be zero. I have called this the *mathematical* postulate of relativity of inertia.

### *3.2 The world matter as an ether in the general theory of relativity*

The necessity of including a world matter in the theory of general relativity is stressed by Einstein in in the address he gave in Leyden [5] in 1920, in which the world matter is rather called an *ether*:

> Recapitulating, we may say that according to the general theory of relativity space is endowed with physical qualities; in this sense, therefore, there exists an ether. According to the general theory of relativity space without ether is unthinkable; for in such space there not only would be no propagation of light, but also no possibility of existence for standards of space and time (measuring-rods and clocks), nor therefore any space-time intervals in the physical sense. But this ether may not be thought of as endowed with the quality characteristic of ponderable media, as consisting of parts which may be tracked through time. The idea of motion may not be applied to it.

---

[4] Actually, thanks to a remark by Cormac O'Raifeartaich (see footnote 1), one should note that Einstein eventually conceded that the de Sitter model did not contain a singularity

In this quotation we can see that the role of the principle of the relativity of inertia is foundational not only on a conceptual ground ("the general theory of relativity space without ether in unthinkable) but also on a metrological ground ("no possibility of existence for standards of space and time").

### *3.3 The abandonment by Einstein of the cosmological term*

The severe criticism made by Eddington to the cosmological model of Einstein based on the instability of the equilibrium induced by the cosmological term, and the remark made by Friedman that the general theory of relativity implies a dynamical rather than a static universe began to lead Einstein to give up his model based on the cosmological term. Finally, when, in 1929 was discovered by Hubble, two years after the theoretical prediction made by Lemaître of the expansion of the universe, Einstein abandoned the constant $\Lambda$, qualifying it as his "biggest blunder".

### *3.4 The principle of the relativity of inertia in modern cosmology*

Now, it turns out that in modern cosmology based on the expansion of the universe and on the quantum physical description of matter, the issues raised both by Einstein and de Sitter can be addressed, and that "system A" and "system B" solutions can be reconciled in an inflationary cosmology such as $\Lambda$CDM (for Lambda Cold Dark Matter, which is the name of the current standard model of cosmology) provided that the two "dark" components of the content of the universe (dark energy and dark matter) can be assimilated to two components of the de Sitter's world-matter, related to the vacuum:

- Because of expansion, the part of the universe that is visible to us, and not the whole universe, is spatially finite: it is a sphere of radius equal to the inverse of the Hubble constant (multiplied by $c$). This boundary of the visible universe is not a singularity, it is a *horizon*

- Although, as noted by de Sitter in the second paper of [4] "In fact, there is no essential difference between the nature of ordinary gravitating matter and the world-matter. Ordinary matter, the sun, stars, etc., are only condensed world-matter, and it is possible, though not necessary, to assume all world-matter to be so condensed", *darkness, namely the absence of non-gravitational interactions,* allows distinguishing observationally world-matter from ordinary matter.

- In a description of non-gravitational interactions of matter based on quantum field theory, the quantum vacuum, namely the ground state of the system of interacting quantum fields with the vanishing of all the occupation numbers, is not the nothingness and can allow to model the world-matter necessary to add to the known visible matter to satisfy the material principle of the relativity of inertia.

The starting point of modern cosmology considering the expansion of the universe and the possible existence of a cosmological constant is the Einstein's equation, which, following the definitions, conventions (the vacuum velocity of light is put to 1), and notations of reference [6] and the proposal of Gliner [7] and Zeldovich [8] to take the cosmological term to the right-hand side[5], reads:

$$\mathcal{R}_{\mu\nu} - \frac{1}{2} g_{\mu\nu} \mathcal{R} = 8\pi G T_{\mu\nu} + \Lambda g_{\mu\nu}. \tag{1}$$

where $\mathcal{R}_{\mu\nu}$ is the Ricci curvature tensor, $\mathcal{R}$ is the scalar curvature, $g_{\mu\nu}$ is the metric tensor, $\Lambda$ is the cosmological constant, $G$ is Newton's gravitational constant, and $T_{\mu\nu}$ is the stress–energy tensor.

The Robertson-Walker metric allows to describe a homogenous and isotropic universe compatible with the Einstein's equation in terms of two cosmological parameters: the spatial curvature index $k$, an integer equal to -1, 0 or +1 and the overall dimensional expansion (or contraction) radius of the universe $R(t)$, depending only on time; note that due to the homogeneity, the geometry actually does not depend on the radial relative coordinate $r$ which is dimensionless:

$$ds^2 = dt^2 - R^2(t)\left[\frac{dr^2}{1-kr^2} + r^2\left(d\theta^2 + \sin^2\theta d\phi^2\right)\right] \tag{2}$$

One often uses a dimensionless scale factor $a(t) = R(t)/R_0$ where $R_0 \equiv R(t_0)$ is the radius at present-day.

In order to derive the Friedman-Lemaitre equations of motion, one assumes that the matter content of the universe is a perfect fluid for which the energy momentum tensor is expressed in terms of the isotropic pressure $P$, the energy density $\rho$, the space time metric $g_{\mu\nu}$ and of the velocity vector $u = (1,0,0,0)$ for the isotropic fluid in co-moving coordinates

$$T_{\mu\nu} = -P g_{\mu\nu} + (P+\rho) u_\mu u_\nu. \tag{3}$$

Since, in all existent homogeneous and isotropic cosmologies, the metric appearing in (1) and in (3) is conformally flat [9], namely proportional to the flat Minkowski metric $\eta_{\mu\nu}$ with a coefficient depending on time, the cosmological term, taken to the right-hand side of the Einstein's equation, can be interpreted as an effective energy-momentum tensor for the vacuum possibly playing the role of a world-matter, rather than as a cosmological constant appearing in the action of the theory. If, as in Eq.(3), one associates a perfect fluid to the cosmological term, its pressure $P_\Lambda(x)$ and energy density $\rho_\Lambda(x)$ sum to zero and the world-matter energy tensor of the vacuum reduces to the pressure term.

---

[5] Actually, thanks to a remark by Cormac O'Raifeartaigh (see footnotes 1 and 4) it may be useful to note that Schrödinger was the first to suggest putting Lambda in the rhs.

## 4/ ΛCDM and the comeback of Lambda

### 4.1 From the Hot Big Bang Model to the new standard model of cosmology

The review by Debono and Smoot[10], in which one can find pedagogical explanations and useful references, shows how theories and observations motivated the adaptation of cosmological models from the first cosmological standard model, the *Hot Big Bang Model* (HBBM) to the new cosmological standard model called the *concordance cosmology* or *Lambda Cold Dark Matter model* (ΛCDM).

The successes of the HBBM include

- the verification of the Hubble Law on the recession of distant galaxies, established through the measurement of their speeds (Doppler effect) and of their distances (by means of the Cepheids)
- the relative abundance of light elements explained thanks to the theory of primordial nucleosynthesis
- the observation of the cosmic microwave background radiation (CMB) at about 3 K, predicted by Gamow and detected, by chance, in 1965 by Penzias and Wilson, then measured with more and more precision (COBE, WMAP, Planck)
- the pure and simple abandonment of the cosmological constant and of all cosmological models denying the expansion of the universe (stationary universe, continuous creation of matter, "tired light", ...)

However, this model suffered from two problems linked to the primordial universe, the horizon problem and the flatness problem.

The horizon problem stems from the fact that, in the Hot Bib Bang Model, the primordial universe, is, as Gabriele Veneziano says, "too large for its age": as a singularity, the big bang implies an expansion of space that is so large that different regions of the universe that can be observed today could not have been in causal contact in the primordial universe; now it turns out that all the CMB regions seem to be in thermal equilibrium.

The flatness problem is known as a "fine tuning problem": already in the sixties, the observations suggested that the density of the total content of the present-day universe is compatible with the so called critical density, namely the one corresponding to the vanishing of the spatial curvature index *k*, a time independent index; but maintaining such a vanishing throughout the history of the universe implies a huge fine tuning that prevents any reliable cosmological modelling.

## *4.2 The assets of ΛCDM*
### 4.2.1 The Friedman-Lemaitre equations

In terms of the pressure and the density of the perfect fluid describing matter, the Friedman equations, with a non-vanishing CC possibly playing the role of a world matter, reads, in terms of the Hubble constant $H = \dot{R}/R$,

$$H^2 \equiv \left(\frac{\dot{R}}{R}\right)^2 = \frac{8\pi G \rho}{3} - \frac{k}{R^2} + \frac{\Lambda}{3} \tag{4}$$

and

$$\frac{\ddot{R}}{R} = \frac{\Lambda}{3} - \frac{4\pi G}{3}(\rho + 3P). \tag{5}$$

Energy conservation leads to a third equation:

$$\dot{\rho} = -3H(\rho + P). \tag{6}$$

### 4.2.2 The three stages of cosmic evolution in ΛCDM

The phases of the cosmic evolution are well represented on the thick line in figure 1 in which the Hubble radius $L = H^{-1}$, the inverse of the time dependent Hubble parameter, is plotted versus the scale parameter *a(t)* (set to 1 today) in logarithmic scale, in such a way that the value zero of the scale parameter *a* (which would lead to a singularity as in the Hot Big Bang Model) is sent to minus infinity. On this figure, one can distinguish three stages[7]:

- The first stage of the evolution is the *primordial inflation stage*, namely a first de Sitter stage occurring at an energy of about $10^{16}$ GeV, during which the Hubble radius is constant (about $10^3$ Planck lengths) whereas the scale factor grows exponentially from α to β in figure 1 by about thirty orders of magnitude.
- At point β, after the end of the primordial inflation, is supposed to have occurred what is usually called the *reheating* of the universe. Point β, when the radius of expansion R(t) reaches about $10^{30}$ Planck's lengths corresponding to the energy scale of CC or to the Compton wave length of a particle with a mass of about 2.5 meV, is the starting point of the second stage of ΛCDM, an expansion phase such as the one in a Hot Big Bang Model, in which point β plays the role of a physical (non-singular big bang), and can be called the *Big Bang ignition point* [11]. In the expansion phase, during which the content of the universe obeys the standard Friedman-Lemaître cosmological equations of evolution, with a time dependence of the cosmic scale $a(t)$ determined by the equation of state parameter $w = P/\rho$ of the component that dominates the evolution at a given epoch, namely;

---

[7] According to a remark by Cormac O'Raifeartaigh (see footnotes 1, 4, 5) one should mention that Lemaître anticipated the three-speed model in the 1930s

- o   An epoch of dominance by radiation $(w=1/3)$ from point β to point ε in figure 1 in which $L \propto a^2$ followed by,

- o   An epoch of dominance by pressure-less matter $(w=0)$ from point ε to point ψ (i.e. today[8]) in which $L \propto a^{1.5}$ ;

- In the third stage, extending from point ψ to point ω in figure 1, the universe will be dominated by the cosmological constant Λ. This stage, like the first one (from point α to point β) is an inflation one (a second de Sitter stage with a scale factor growing exponentially with the cosmic time, and a Hubble radius slowly increasing asymptotically to $\sqrt{3/\Lambda}$ ) called the *late inflation stage* characterized by an equation of state $w = -1$ compatible with the present day observation, $\left(w = -1.019^{+0.075}_{-0.080}\right)$ [20].

### 4.2.3 The flatness sum rule

It is useful to define a density, called the *critical density*

$$\rho_c \equiv \frac{3H^2}{8\pi G} \tag{7}$$

which would be a solution to the Friedman's equation (4) if the curvature index $k$ and CC were zero. With respect to this critical density one defines for each component, including the one of CC, the relative contribution to the critical density, called its cosmological parameter, and rewrite the present day Friedman's equation (4) as

$$\Omega_{tot} = \rho / \rho_c$$
$$k / R^2 = H^2 (\Omega_{tot} - 1) \quad (8)$$
$$k / R_0^2 = H_0^2 (\Omega_M + \Omega_R + \Omega_\Lambda - 1)$$

where the subscript M stands for pressureless matter, (or "dust fluid"), the subscript R stands for radiation (or relativistic particles) and $\Omega_\Lambda = \Lambda / 3H^2$. Since the curvature index $k$ does not depend on time, its vanishing at present day implies its vanishing at all epochs, so in terms of time-dependent densities, the Friedman's equation (4) takes the form of the *flatness sum rule* which, in terms of densities reads

$$\rho_M + \rho_R + \rho_\Lambda = \rho_c \quad (9)$$

---

[8]Actually, due to the compressing of time scales in the cosmic evolution in the late epoch, the point ψ in figure 1 does not correspond to 'today', but rather to the transition from expansion to "re-inflation" $(\rho + 3P = 0)$; the point corresponding to 'today' would be slightly on the right of ψ.

### 4.3 The emergent perspective of gravity and the first stage of the comeback of $\Lambda$: an integration constant in the solution of the Friedman Lemaître equation

It is tempting to interpret the flatness sum rule as expressing the vanishing of the total (gravitational plus kinetic) energy density, equal to $\rho_c$, to which has to be subtracted the vacuum energy density $\rho_\Lambda$, thus qualifying ΛCDM as a cosmology of emergence out of the vacuum, a so-called "free lunch cosmology". Actually, such an interpretation is suggested in [6] in the comment made about the Friedman's equation (4): "By interpreting $-k/R^2$ Newtonianly as a 'total energy', we see that the evolution of the Universe is governed by a competition between the potential energy, $8\pi G\rho/3$, and the kinetic term $(\dot{R}/R)^2$". But, this suggestion is criticized a few lines below in the following way: "Note that the quantity $-k/R_0^2 H_0^2$ is sometimes referred to as $\Omega_k$. This usage is unfortunate: it encourages one to think of curvature as a contribution to the energy density of the Universe, which is not correct."

However, I think that such an interpretation of the Friedman's equation can be made correct, if, as advocated by Padmanabhan [12] and as I am going to explain now, one adopts the *emergent perspective of gravitation* per which the quantity that is conserved in the cosmic evolution is not a 'total energy' but rather a thermodynamic potential (i.e. defined up to an arbitrary additive constant), namely an *enthalpy* or total *heat* content.

The idea underlying the emergent perspective of gravitation is that in general relativity, *horizons* are unavoidable, and that, since horizons block information, *entropy* can be associated, through them, to space-time, and thus that space-time *has a micro-structure*. The Davies-Unruh [13] effect, the thermodynamics of black holes of Hawking [14] and Bekenstein [15], the Jacobson [16] interpretation of the Einstein's equation as an equation of state, or the interpretation of gravity as an entropic force by Verlinde ([17] and more recently [18]) say, rely on this idea which provides a possible thermodynamic route toward quantum gravity.

In the conventional approach, gravity is treated as a field which couples to the energy density of matter. The addition of a cosmological constant – or equivalently, shifting of the zero level of the energy – is not a symmetry of the theory since the gravitation field equations (and their solutions) change under such a shift. But in the EPG, rather than the *energy density* it is the *entropy density* which plays the crucial role and shifting the zero level of the entropy is now a symmetry of the theory.

In ΛCDM, the time-dependent null surface, with radius $H^{-1}$ blocks information and can thus be endowed with an entropy [15] proportional to its area

$$S = \left(A/4L_P^2\right) = \left(\pi/H^2 L_P^2\right) \tag{10}$$

where $L_P = \left(\hbar G / c^3\right)^{1/2}$ is the Planck's length, and a temperature [14]

$$T = \hbar H / 2\pi \tag{11}$$

During a time-interval *dt*, the change of the gravitational entropy (i.e. the entropy associated with space-time) is

$$\left(dS/dt\right) = \left(1/4L_P^2\right)\left(dA/dt\right) \tag{12}$$

and the corresponding *heat* flux

$$T\left(dS/dt\right) = \left(H/8\pi G\right)\left(dA/dt\right) \tag{13}$$

For the matter contained in the Hubble volume, the classical (Gibbs-Duhem) thermodynamic relation tells us that the entropy density is $s_m = (1/T)(\rho + P)$, corresponding to a heat flux through the boundary equal to

$$TS_m A = (\rho + P)A \tag{14}$$

Balancing gravitational (12) and matter (14) heat flux equations leads to $\dfrac{H}{8\pi G}\dfrac{dA}{dt} = (\rho + P)A$, which, with $A = 4\pi/H^2$ gives

$$\dot{H} = -4\pi G(\rho + P) \tag{15}$$

Now, energy conservation for matter leads to

$$\begin{aligned}\frac{d(\rho a^3)}{dt} &= -P\frac{da^3}{dt} \\ \dot{\rho} &= -3H(\rho + P)\end{aligned} \tag{16}$$

which is nothing but eq. (6) and which, combined with eq. (15) and integrating over time leads to

$$\frac{3H^2}{8\pi G_N} \equiv \rho_c = \rho + \text{ arbitrary constant} \tag{17}$$

Comparing this last equation with Eq.(9), one sees that since the entropy density vanishes for the cosmological constant, the arbitrary constant can be put to $\rho_\Lambda$ that acts as an integration constant because $\rho_c \xrightarrow[t \to \infty]{} \rho_\Lambda$. This feature is the main interest of the emergent perspective of gravitation: instead of being a parameter in the action of the theory, the cosmological constant appears as an integration constant in a particular solution of the equation. Now, as Padmanabhan says in the conclusion of his textbook on gravitation [19]

> "The integration constants which appear in a particular solution have a completely different conceptual status compared with the parameters that appear in the action describing the theory. It is much less troublesome to choose a fine-tuned value for a particular integration constant in the theory if observations require us to do so. From this point of view, the cosmological constant problem is considerably less severe when we view from gravity from the emergent perspective"

Eq. (9) thus becomes

$$\rho_M + \rho_R = \rho_c - \rho_\Lambda = \Lambda_{\text{eff}} / 8\pi G_N \tag{18}$$

Where the left-hand side is the sum of energy densities of all the components of the universe (baryonic, relativistic and dark matters) contributing to gravitation, whereas the right-hand side equated to an *effective cosmological constant* $\Lambda_{\text{eff}}$, can be interpreted as the energy density of the world-matter corresponding to a negative pressure and contributing to inertia, just as the "constant Λ" term in the system A or system B of Einstein and de Sitter expresses the principle of equivalence of gravitation and inertia, so Eq. (18) can be re-written as

$$\rho(\text{Matter}) + P(\text{World-Matter}) = 0 \tag{19}$$

### *4.4 The primordial inflation and the second stage of the comeback of Λ: as a fifth foundational constant.*

The primordial inflation phase cures the defects of the simple big bang model implied by the existence of a singularity. During the primordial inflation phase, the "young universe" is no more "too large for its age", the whole content of the universe is in causal contact, which solves the horizon problem. Inflation also solves the flatness problem: it flattens space in such a way that we can assume that at its end the spatial curvature is already compatible with zero, in agreement with the present-day observations $\left(\Omega_k \equiv -k/R_0^2 H_0^2 = 0.0008^{+0.0040}_{-0.0039}\right)$ [20].

Although this primordial inflation process remains conjectural (is it induced by a new ad hoc quantum field, the "inflaton" or strictly related to the metric field?), the main asset of ΛCDM is to have provided inflation with the credibility it was lacking beforehand.

The fact that the scale at which this primordial inflation is compatible with the one at which Grand Unified Theories (GUT) are expected to be at work suggests that it could also allow solving a third problem of the HBBM, the one of the absence of magnetic monopoles that would be predicted to be produced in large numbers in the GUT framework: inflation would expel the GUT magnetic monopoles beyond the horizon. More generally, there is a wide consensus to consider the primordial inflation as belonging to the realm of all the issues that are beyond the standard models of elementary particles and of cosmology (BSM), including the ones related to pre-geometric or quantum gravity, namely gravity at the Planck's scales of length $L_P = \sqrt{\hbar G/c^3}$, time $T_P = \sqrt{\hbar G/c^5}$ and energy $E_P = \sqrt{\hbar c^5/G}$.

It is remarkable that the primordial inflation ends when the radius of expansion $R(t)$ reaches about $10^{30}$ Planck's lengths corresponding to an energy scale that is quite compatible with the one of Λ or to the Compton wave length of a particle with a mass of about 2.5 meV, which strongly suggests that the domain in which one can understand the true meaning of Λ is the one of BSM physics, including the realm of quantum gravity. Now it turns out that such a connection of Λ with quantum gravity can be guessed in the emergent perspective of gravity (EPG) discussed above. According to EPG, which assumes that spacetime has a microstructure in terms of "spacetime atoms", the entropy of the Hubble horizon determined by CC is proportional to its area $1/\Lambda$ expressed in the Planck's unit of area $A_P = \hbar G/c^3$. This means that the area of the Hubble horizon divided by the Planck's area is proportional to the "*Avogadro constant of spacetime*", $N_A^{ST}$ which, when multiplied by the Boltzmann's constant $k$, represents to the total (very large, but *finite*) content of relevant information in the observable universe *cosmIn*.

$$\frac{c^3 k}{\hbar G \Lambda} = \text{cosmIn} = N_A^{ST} k \tag{20}$$

In [21], T. and H. Padmanabhan show that (i) the numerical value of the cosmological constant, as well as (ii) the amplitude of the primordial, scale invariant, quantum fluctuation spectrum can be determined in terms of a single free parameter, $a_{QG}$ which specifies the energy scale at which the universe makes a transition from a pre-geometric or quantum phase to the classical phase. For a specific value of the parameter, they obtain the correct results for both (i) and (ii).

## 5/ The foundational role played by the five constants considered by Einstein

### *5.1 "Clocks and measuring rods", the metrological concern of Einstein*

In several circumstances, Einstein expressed his concerns about clocks and measuring rods, about which he had to make some unavoidable, based on experience, assumptions. For instance, in one of his lectures given at Princeton in 1920, he says [22]:

> "In this the physical assumption is essential that the relative lengths of two measuring rods and the relative rates of two clocks are independent, in principle, of their previous history. But this assumption is certainly warranted by experience; if it did not hold there could be no sharp spectral lines; for the single atoms of the same element certainly do not have the same history, and it would be absurd to suppose any relative difference in the structure of the single atoms due to their previous history if the mass and frequencies of the single atoms of the same element were always the same."

In retrospect, it seems that the point of view of Einstein about the universal constants is not far from the one adopted in the current methodology of redefinition of the SI. This can be seen in the following quotation of the scientific autobiography of Einstein [23]:

> "The speed of light $c$ is a quantity which intervenes as a 'universal constant' in the equations of physics. But if one takes as a unit of time, not the second, but the time that light takes to go 1 cm, $c$ no longer appears in the equations. In this sense, the constant c is only an apparent universal constant. It is manifest, and universally admitted, that it would also be possible to eliminate universal constants by introducing instead of the gram and the centimeter, adequately selected 'natural' units (e.g. mass and the radius of the electron). (…) Imagine that this has been realized [elimination of two universal constants to the benefit, for example of the mass and the radius of the electron]; then there appear in the fundamental equations of physics only dimensionless constants. About them, I would like to enunciate a principle which, provisionally, cannot be based on nothing else than on my confidence in the simplicity, or rather in *the intelligibility of Nature*: there is no arbitrary constant of this type. In other words: Nature is such that it is logically possible to establish laws that are so strongly defined that only constants susceptible of a complete rational determination appear in them (there are therefore no constants whose numerical values can be modified without the theory being destroyed)"

The same idea is also put forward in the following sentences from a 1935 article "Physics and reality [24]"

> "The very fact that the totality of our sense experiences is such that by means of thinking (operations with concepts, and the creation and use of definite functional relations between them, and the coordination of sense experiences to these concepts) it can be put in order, this fact is one which leaves us in awe, but which we shall never understand. One may say 'the eternal mystery of the world is its comprehensibility.' It is one of the great realizations of Immanuel Kant that the setting up of a real external world would be senseless without this comprehensibility."

### *5.2 The incompleteness reproach made to "quantum statistics" by Einstein*

It is in the correspondence he exchanged with Schrödinger, a few days after the publication of the famous EPR paper [25] that Einstein expressed in the clearest way his criticism on quantum physics [26]:

> "From the point of view of the principles, I absolutely do not believe in the existence of a statistical basis of physics, in the sense that quantum mechanics understands - and this, despite the successes that this formalism has gained case by case (...). I find that giving up a spatio-temporal apprehension of the reality is an idealistic and spiritualistic position.

And he formulated the requirement that any description of reality should satisfy to be qualified as "complete"

> In quantum theory, a real state of a system is described by a normed function of the coordinates $\psi$ (of the configuration space). The evolution over time is unequivocally given by the Schrödinger equation. We would like to be able to say: $\psi$ is in a one-to-one coordination with the actual state of the real system. The statistical character of the measurement results is exclusively to be due to the measuring devices or the measuring procedures. When it works, I speak of a complete description of reality by theory. But, if such an interpretation proves impracticable, I say that the theoretical description is "incomplete".

### *5.3 The informational turn: the recognition of the foundational role of information*

Until recently, the consensus about the meaning of the universal constants, was that three constants, and only three, the gravitational constant, the speed of light, and the Planck constant, determine both the fundamental laws and the fundamental units of the general framework of physics. Considering these constants, one by one, two by two, or all three together, leads, according to this consensus, to structuring physics into subdisciplines that we can symbolically represent as the summits of a cube, "the cube of theories", What I call the *informational turn of physics* is the recognition of the foundational role of information and, thus, the inclusion of the Boltzmann's constant $k$ that has the dimensional content of an information (or an entropy) in the panoply of the foundational constants.

The informational turn is well explicated by Alexei Grinbaum in his paper [27] *On Epistemological Modesty*, in which he quotes another excerpt of the scientific autobiography of Einstein

> "One is struck [by the fact] that the theory [of special relativity] introduces two kinds of physical things, i.e., (1) measuring rods and clocks, (2) all other things, e.g. the electromagnetic field, the material point, etc. This, in a certain sense, is inconsistent; strictly speaking measuring rods and clocks would have to be represented as solutions of the basic equations (objects consisting of moving atomic configurations), not, as it were, as theoretically self-sufficient entities. However, the procedure justifies itself because it was clear from the very beginning that the postulates of the theory are not strong enough to deduce from them sufficiently complete equations in order to base upon such a foundation a theory of measuring rods and clocks." (Einstein, 1969, p. 59)

> Epistemologically, it is unreasonable to expect, as Einstein did, that the theory of measuring rods and clocks could be based on a set of yet stronger postulates that would, at the same time, provide also an account of all physical phenomena measured by means of these rods and clocks. To see why Einstein found himself at an impasse, albeit an unnecessary one, consider the following schematic representation of physical theories. Assume that phenomena are best described by theories that are interconnected in the form of loop. Any particular theory is represented by cutting the loop at some point and thus separating the target object of the theory from the theory's presuppositions. (…) Consider the loop between physical theory and information. Physics and information mutually constrain each other, and every theory will give an account of but a part of the loop, leaving the other part for metatheoretic assumptions. In a first loop cut, information lies in the meta-theory of the physical theory, and physics is therefore based on information. In a different loop cut, informational agents are physical beings, and one can describe their storage of, and operation with, information, by means of effective theories that are reduced, or reducible in principle, to physical theory.

The way out of the impasse in which Einstein was stuck is not to adopt the cutting of the epistemic loop that leads to the statement that "information is physical" but rather the one leading to the statement that "physics is informational ". Now this is precisely what Jeffrey Bub proposes to do in his paper [28] untitled *Why the quantum?*

> i) A quantum theory is best understood as a theory about the possibilities and impossibilities of information transfer, as opposed to a theory about the mechanics of nonclassical waves or particles.
>
> ii) Given the information-theoretic constraints, any mechanical theory of quantum phenomena that includes an account of the measuring instruments that reveal these phenomena must be empirically equivalent to a quantum theory.
>
> iii) Assuming the information-theoretic constraints are in fact satisfied in our world, no mechanical theory of quantum phenomena that includes an account of measurement interactions can be acceptable, and the appropriate aim of physics at the fundamental level then becomes the representation and manipulation of information.

The informational turn is also what, according to Anton Zeilinger, provides quantum physics with the "generally accepted conceptual foundation" it was missing in contrast with the general theory of relativity;

> "In contrast to the theories of relativity, quantum mechanics is not yet based on a generally accepted conceptual foundation. It is proposed here that the missing principle may be identified through the observation that all knowledge in physics has to be expressed in propositions and that therefore the most elementary system represents the truth value of one proposition, i.e., it carries just one bit of information. Therefore, an elementary system can only give a definite result in one specific measurement. The irreducible randomness in other measurements is then a necessary consequence. (…) The universe is participatory at least in the sense that the experimentalist by choosing the measurement apparatus, defines out a set of mutually complementary observables which possible property of a system can manifest itself as reality and the randomness of individual events stems from the *finiteness of information* [underlined by me]. (…) In conclusion, it may be very well said that information is the irreducible kernel from which everything else flows. Then the question why nature appears quantized is simply a consequence of the fact that information itself is quantized by necessity."

## *5.4 The informational turn and the answer to the incompleteness reproach made by Einstein to "quantum statistics"*

Schematically the overall outcome of the informational turn can be shown by the replacement of the theoretical tripod of twentieth century physics (see figure 2) by the new theoretical tripod of figure 3, in which "quantum statistics" is replaced by "quantum information theory" (QIT), and "General Relativity" is replaced by "quantum spacetime theory" (QSTT). These replacements mean that the three foundational theories of the new tripod which each, involve an elementary quantum (the quantum of action in QFT, a quantum of spacetime area $A_P = \hbar G / c^4$ in QSTT and a quantum of information in QIT), do not imply any recourse to statistics. This recourse is confined in the intersection of the QFT and QIT domains, which is the realm of quantum mechanics and of the high energy standard model (HEPSM) and, on the other hand, in the intersection of the QIT and QSTT domains, which is the realm of astrophysics and of the cosmological model (CSM). As belonging to the phenomenological realm, these two standard models use statistical methods in the comparison with experiment or observation.

About the interpretation of quantum physics, the result of the informational turn is that the complementarity should be considered as that between the means of observation pertaining to the quantum theory of information and the observed system pertaining to quantum mechanics or to quantum field theory, rather than that between two classically contradictory limits (waves or corpuscles, for example) of quantum physics. According to the proposed new interpretation, *matter/information complementarity* is to be considered as the result of a *double quantization* involving the quantum of action and the quantum of information simultaneously, expressed by the key concept of *probability amplitude*, of which the modulus refers to the probability, i.e. the information and the phase to the coherence, i.e. the action.

In cosmology, it has been realized, thanks to the informational turn, that the entropy that is involved in the thermodynamics of spacetime underlying the emergent perspective of gravitation is not a statistical entropy but rather an entanglement entropy in such a way that it does not imply the existence of any hidden degree of freedom [29]. On an interpretive ground, the cosmological standard model relies on *spacetime/information complementarity* involving the spacetime quantum and information quantum

In the third intersection, the one of the QFT and GSTT domains, the investigations about physics lying beyond the standard models (BSM) can ignore statistics as far as they remain as purely theoretical conjectures about an ideal physics at zero temperature.

The intersection of the three domains of QFT, QIT and QSTT is the domain of the purely theoretical program of building a supposedly complete *thrice quantized gravity theory* (3QGT), considering the five universal constants assembled in *cosmIn* – see Eq. (20).

## 6/ Conclusion

"Nature is earlier than man, but man is earlier than natural science", this aphorism of Von Weiszäcker, quoted by Zeilinger to strengthen his argument in favor of the informational turn, could also have been used by Michel Bitbol in favor of the *reflective turn* [30] in the philosophy (metaphysics) of quantum physics that he proposes:

> "Instead of either formulating new metaphysical images (as realists would do) or rejecting any metaphysical attempt (as empiricists would do), the case of quantum mechanics might well require from us a complete redefinition of the nature and task of metaphysics. The sought redefinition can be performed in the spirit of Kant, according to whom metaphysics is the discipline of the boundaries of human knowledge. This can be called a "reflective" conception of metaphysics. In this paper, each one of the most popular "interpretations" of quantum mechanics is shown to be naturally associated with a variety of Kant-like reflective metaphysics. Then, the two major "paradoxes" of quantum mechanics (the measurement problem and the EPR correlations) are reformulated by way of this reflective attitude, and they are thereby "dissolved". Along with this perspective, quantum mechanics becomes one of the most elegant and understandable theories of the history of physics in addition of being one of the most efficient. The only point that must be clarified is why it looks culturally so difficult to accept a reflective and non-ontological standpoint on physical theories."

In conclusion, it must be stressed that this reflective turn is also, as said above, what underlies the redefinition of the SI because the five universal constants *c, h, k, G and Λ* on which it is based are all related to the boundaries of objective knowledge provided by physics.

**]Acknowledgements** It is a pleasure to acknowledge Martin Milton, Estefania de Mirandès and Janet Miles of the BIPM for their invitation to the Symposium on fundamental constants held in September 2017**,** and Cormac O'Raifeartaigh who signaled to me the history of one hundred years of the cosmological constant [31]

Figures

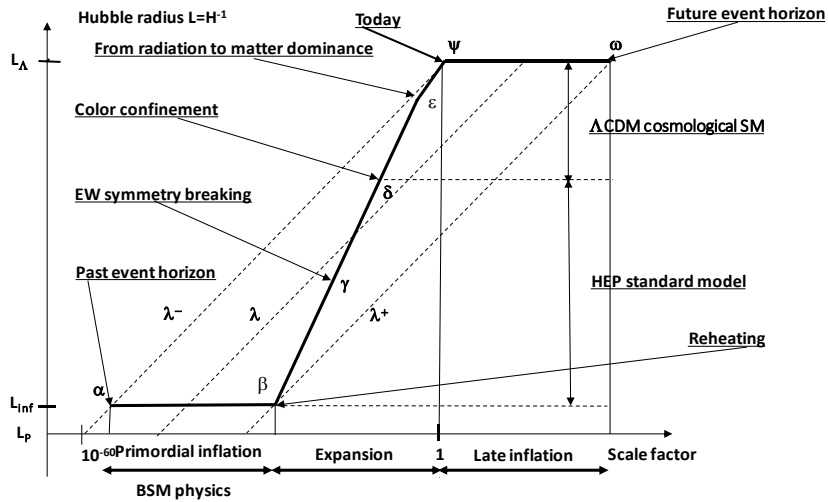

Figure 1: The three stages of ΛCDM

**Caption** The ΛCDM cosmology represented in a graphic in which the Hubble radius $L = H^{-1}$ is plotted versus the scale factor $a(t)$ (set to 1 today) in logarithmic scale. The cosmic evolution is schematized on the thick line, on which the cosmic time grows linearly in the inflation phases (horizontal parts from point α to β and from ψ to ω) and exponentially in the expansion phase (from β to ψ). All quantum fluctuations with generic wave-length λ exit from the Hubble horizon in the primordial inflation phase enter it in the expansion phase, and re-exit it in the late inflation phase. No information carrying quantum fluctuation with a wave-length smaller than $\lambda^-$ or larger than $\lambda^+$ enters the Hubble horizon. From this limitation, one infers that the total content of information in the observable universe, which is proportional to inverse of the cosmological constant is **finite**.

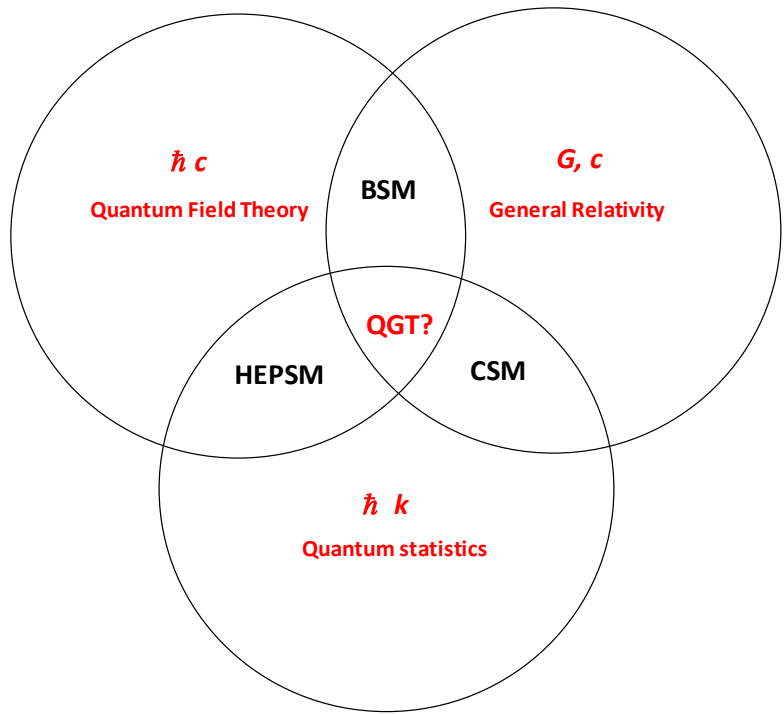

Figure 2 The theoretical/phenomenological landscape at the end on the 20[th] century

**Caption** BSM = Beyond the Standard Models; QGT? = Quantum Gravity Theory (the question mark means that this theory is yet to be discovered); HEPSM = High Energy Physics; CSM = Comological Standard Model

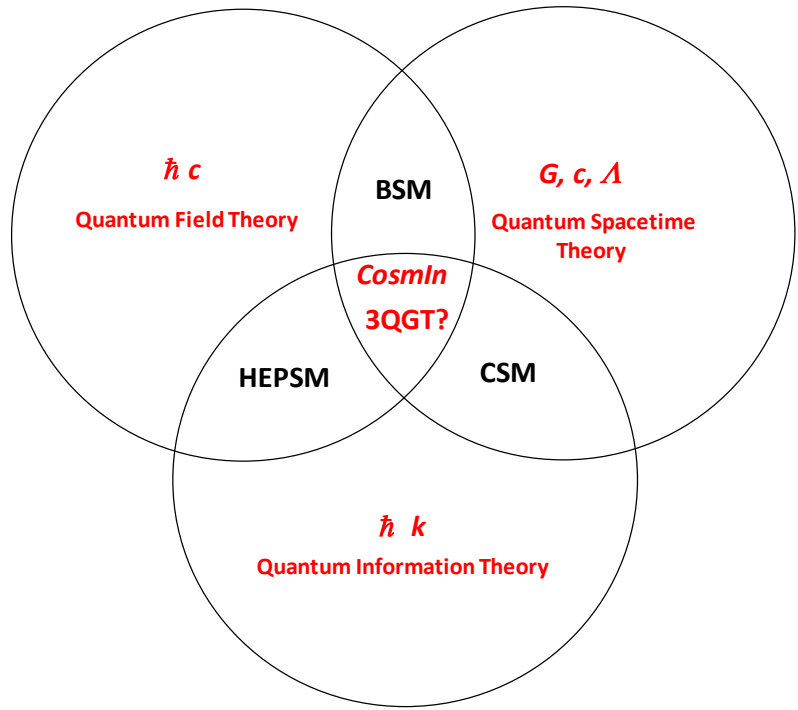

Figure 3 Towards a new Theoretical/phenomenological Landscape

**Caption** With respect to figure 2, "Quantum Information Theory" has replaced "Quantum Statistics"; 3QGT (Thrice Quantized Gravity Theory) has replaced QGT; mention is made to "CosmIn", the total content of information in the observable universe – see Eq. (20) and the caption of figure 1.